\documentclass[runningheads]{llncs}
\usepackage[T1]{fontenc}
\usepackage{graphicx}
\usepackage{enumitem}
\usepackage{hyperref}
\usepackage{listings}
\usepackage{amsmath}
\usepackage{twemojis}
\usepackage{tcolorbox}
\usepackage{mathabx}
\usepackage{csquotes}

\definecolor{RoyalBlue}{RGB}{65, 105, 225}
\newcommand{\tellxr}[1]{%
  {\footnotesize \texttwemoji{robot}   \textsf{\color{RoyalBlue} \raggedright #1} \newline }
}

\newcommand{\tellxrInline}[1]{%
  {\footnotesize \texttwemoji{robot}   \textsf{\color{RoyalBlue} \raggedright #1}  }
}

\definecolor{MediumPurple}{RGB}{147, 112, 219}
\definecolor{Crimson}{RGB}{220, 20, 60}
\newcommand{\angie}[1]{%
  {\footnotesize \texttwemoji{woman: light skin tone}  \textsf{\color{Crimson} \raggedright #1} \newline }
}

\newcommand{\angieInline}[1]{%
  {\footnotesize \texttwemoji{woman: light skin tone}  \textsf{\color{Crimson} \raggedright #1} }
}

\definecolor{ForestGreen}{RGB}{34, 139, 34}
\newcommand{\bob}[1]{%
  {\footnotesize \texttwemoji{man: light skin tone} \textsf{\color{ForestGreen} \raggedright #1} \newline }
}

\newcommand{\bobInline}[1]{%
  {\footnotesize \texttwemoji{man: light skin tone} \textsf{\color{ForestGreen} \raggedright #1} }
}

\begin{document}
%
\title{Tell-XR: Conversational End-User Development of XR Automations \thanks{This is a preprint of a manuscript submitted for consideration in LNCS, Springer.}}
\titlerunning{Tell-XR: Conversational EUD for XR}
%

\author{Alessandro Carcangiu\inst{1} \and Marco Manca\inst{2} \and Jacopo Mereu\inst{1} 
        \and Carmen Santoro\inst{2} \and Ludovica Simeoli\inst{2} \and Lucio Davide Spano\inst{1}}

\institute{University of Cagliari, Dept. of Mathematics and Computer Science, Via Ospedale 72, 09124, Cagliari, Italy \email{\{alessandro.carcangiu, jacopo.mereu, davide.spano\}@unica.it} \and 
ISTI-CNR, HIIS Laboratory, Via G. Moruzzi 1, 56124, Pisa \email{\{marco.manca, carmen.santoro, ludovica.simeoli\}@isti.cnr.it}}

\authorrunning{A. Carcangiu et al.}
%
%
\maketitle              
\begin{abstract}
The availability of extended reality (XR) devices has widened their adoption, yet authoring interactive experiences remains complex for non-programmers. We introduce Tell-XR, an intelligent agent leveraging large language models (LLMs) to guide end-users in defining the interaction in XR settings using automations described as Event-Condition-Action (ECA) rules. Through a formative study, we identified the key conversation stages to define and refine automations, which informed the design of the system architecture. The evaluation study in two scenarios (a VR museum and an AR smart home) demonstrates the effectiveness of Tell-XR across different XR interaction settings. 

\end{abstract}
%
%
%

\newcommand{\Cagliari}{Institute-1 }
\newcommand{\Pisa}{Institute-2 } 

\section{Introduction}

The increasing availability of affordable devices supporting extended reality (XR) has broadened the potential audience for XR experiences.  
However, developing XR content remains expensive due to the required expertise~\cite{statista2024xr}.
%
Recent research has aimed to make XR authoring tools more accessible for prototyping the interaction design~\cite{protoar,leiva2021,nebeling2020xrdirector,360proto,speicher2021}. This approach seeks to involve end users in the later phases of experience development, as they are often more aware of their needs~\cite{zarraonandia2016inmersive,pacpac1,pacpac2,ecarules4all,yigitbas2023}.
However the barriers for people without programming or design experience remain high~\cite{nebeling2023,ashtari2020}. 

In parallel, the Internet of Things (IoT) connects smart objects that interact, exchange data, and respond autonomously to different types of events. These devices influence the physical world by triggering actions and performing services. In the IoT domain, the research demonstrated that end-users could leverage user-friendly tools to control and personalise automation based on device services, enabling a more personalised experience and aligning with their preferences and needs~\cite{gallo2022,gallo2023}. Combining XR-ready smartphones and HMDs together with the potentialities of IoT could push this trend further, by blending information, interactions and services between the virtual and the real world~\cite{heun2013reality,seiger2021,sun2019,mahroo2019}. 

Furthermore, the introduction of Large Language Models (LLM) in the XR authoring is a great opportunity to support end-users. Current approaches range from creating virtual objects from voice commands~\cite{giunchi2024}, generating Unity code for defining behaviour and interactions~\cite{delatorre2024} or as a mean to combine different ideas collaboratively~\cite{kucu2024}. However, one of the main limitations of these approaches is that the output consists of artifacts (e.g., C\# code or 3D models) the end-users has no way to inspect and understand. 

In this paper, we introduce Tell-XR, an intelligent agent based on LLMs able to guide end-users in creating automations in XR described as event-condition-action (ECA) rules. This work i) identifies the key stages for end-users to define and refine automations, ii) proposes a general LLM-based bot architecture using multimodal information for automation in XR, and iii) validates the approach through a user study in a virtual reality (VR) museum and an augmented reality (AR) smart home scenarios.


\section{Related Work}
\textbf{End-User Development for XR}.
Authoring tools for end users in eXtended Reality empower individuals to create, modify, and personalise immersive experiences. Still, they often impose specific processes or design limitations, especially considering the available interactions~\cite{nebeling2018ismar,nebeling2023}.
Creating interactive virtual content involves two primary steps: modelling the virtual shapes and defining their behaviours in response to user interactions or other contents~\cite{lee2004}. 
Nebeling et al.~\cite{nebeling2018ismar} identify authoring platforms having a steep learning curve (such as Unity~\cite{unity} or Unreal~\cite{unreal}), which are too complicated for non-professional users, and those targeting people without programming experience for rapid prototyping of virtual environments~\cite{ashtari2020}. 
The latter covers the case of research tools like Spacetime~\cite{xia2021}, 360proto~\cite{360proto} and 360theater~\cite{speicher2021} where the creation of VR prototypes is simplified through different means. 
An effective solution for defining the interactive behaviour is simulating it using Wizard of Oz techniques~\cite{alce2015wozard,360proto,speicher2021}. In XRDirector~\cite{nebeling2020xrdirector}, collaborative multiuser authoring is facilitated across various devices like phones, VR and MR headsets, and screens. 
Although these approaches require a low entry threshold~\cite{myers2000}, they constrain prototypes to low fidelity. Another way to maintain a low threshold is to limit the expressiveness (i.e., the ceiling).
For instance, VR GREP~\cite{zarraonandia2016inmersive} limits the interaction to selecting elements and navigation in the world, while visual programming immersive interfaces~\cite{ens2017ivy,heun2013reality,wang2020capturar} provide pre-defined behaviour or animation blocks. 
FlowMatic~\cite{zhang2020flowmatic} provides more expressiveness, introducing a flow-based diagram supporting functional reactive programming~\cite{elliott1997functional}. Still, its vocabulary is suited for novice programmers who can understand the node properties and the effect of their connections. We exploit LLMs to support high-fidelity experiences while keeping a low threshold. 

Text-based approaches rely on rules for describing behaviours using constrained natural language. Ble\v{c}i\'{c} et al.~\cite{pacpac1,pacpac2} devised a rule-based authoring tool for developing point-and-click games. Artizzu et al.~\cite{ecarules4all} expanded the approach by supporting end users' customisation of VR environments through a text-based panel while immersed in VR.
XRSpotlight~\cite{xrspotlight} supports the specification of rules that involve different modalities as triggers. However, all these approaches rely on end-user's abstraction abilities to create rules.  

Several End-User Development (EUD) approaches have explored also Augmented Reality (AR) and Mixed Reality (MR) for IoT automation. Reality Editor \cite{heun2013reality} allows users to link physical objects via graphical elements, but lacks support for specifying joint behaviours across multiple objects. HoloFlows \cite{seiger2021} simplifies IoT workflow configuration using a no-code MR interface, but it requires dedicated hardware and is focused on linking nearby devices.
MagicHand~\cite{sun2019} and HoloHome \cite{mahroo2019} enable IoT control via AR and gesture-based interactions but are limited to single-device control. BricklAyeR \cite{stefanidi2019bricklayer} introduces a 3D building block metaphor for defining trigger-action rules, but it was only evaluated through expert walkthroughs. MagiPlay \cite{stefanidi2020} extends this approach as a learning tool for children, though it does not support mobile automation.
ARticulate \cite{clark2022} facilitates function discovery in smart spaces but does not handle multi-object automation. SAC \cite{ariano2023} attempts to support automation control but requires close proximity to devices. 
ProInterAR \cite{ye2024} allows AR scene creation via a tablet and AR-HMD but may face efficiency challenges due to device switching. PRogramAR~\cite{ikeda2024} targets AR-based trigger-action programming for robot control, aiming to assist non-technical users.
All these approaches are strongly tied to a specific domain, or they support a single XR configuration (AR or VR).

\noindent\textbf{LLMs for automations}.
Recent advances in LLMs have led to increased research in generating automations expressed as rules derived from natural language conversations, especially for smart homes \cite{gallo2022,gallo2023}.
However, the challenges of enabling unskilled users to customize their applications in XR using LLMs have received less attention.
In \cite{delatorre2024}, the authors present a framework that leverages LLMs for virtual object generation in 3D scene development, consisting of five modules (planner, scene analyser, skill library, builder and inspector), producing C\# code. 
The approach lacks transparency and control for end-users, as scene editing occurs in a professional development tool (Unity).

MagicItem~\cite{kurai2024} integrates large language models (LLMs) with the scripting capabilities of the Cluster metaverse platform, enabling non-programmers to define object behaviour using natural language. The evaluation with 63 users shows that even individuals without programming experience could successfully create VR object behaviours, highlighting the tool's potential to democratize VR authoring. However, the system assists only in defining basic behaviours, and further research is required on how to specify more complex ones, such as dynamic those occurring only under specific events and/or conditions.

AtomXR~\cite{cai2023}, is a no-code immersive prototyping tool addressing usability issues in Extended Reality (XR) content creation, featuring natural language UIs LLMs, multimodal inputs (such as eye-gaze and touch). 
It exploits a high-level scripting language, AtomScript, allowing users to create content directly in XR, reducing the need for context switching between development and testing environments. Two user studies demonstrated improved prototyping speed and user satisfaction compared to traditional methods. However, the studies involved participants with varied levels of prior programming experience, thereby not specifically focusing on unskilled users; in addition, it is not clear to what extent that system is able to support complex tasks.

\section{Formative Study}
Our approach enables user-driven customization of XR experiences, from VR scenes to AR-based home automation. In order to understand which are the typical patterns the end-users follow in describing their automation intent, we run a formative study in April 2024, using the Wizard of Oz (WoZ) method. One of the researchers impersonated the LLM-based agent (i.e., the bot), interpreting and answering participant's questions and commands, and proposing the rules resulting from the dialogue. The study consisted of two sessions, one considering a VR museum and the other one considering an AR smart home. 
To simulate the virtual museum, we set up a room featuring small statues and images of paintings displayed on panels. Additionally, we incorporated small devices such as buttons, lamps, pedestals, and other interactive elements that could be used for automation. Interactive functions and multimedia content were simulated using paper for text, speakers for audio, and a tablet for video playback.
To simulate a smart home in AR, we equipped a research lab room with various smart devices, including real items (like smart lamps and sensors) and a few simulated ones (like a refrigerator, represented with paper artifacts). Each device had a card indicating its connectivity for automation and additional information provided by a researcher using paper displays. We also set up a multi-purpose panel to simulate services such as weather, location, and time for potential automations.
Each session consisted of the same six scenario-adapted tasks:

\textbf{T1: Exploration}. Participants familiarised with the environment and the simulated VR and AR contents. 

\textbf{T2 and T3: Simple Automation}. Participants talked to the simulated bot for creating an automation consisting of a single event and a single action (trigger-action). In VR the tasks consisted in displaying an information panel when the user points an artwork (T2) and turning on an artwork spotlight when getting close to it (T3). In AR, the participant created an alarm when a sensor detects smoke (T2) and turned on the home entrance light after a certain time at night (T3). 

\textbf{T4 and T5: Complex Rules}. Participants had to create two complex automations (i.e. including conditions) with the help of the simulated bot. In VR the task consisted in showing additional information on one statue only if the visitor already visited a painting (T4) and to start the audio content when the user is close to an artwork and presses a button (T5). In AR, it consisted in turning on the air purifier if the user is at home, and turning it off when not at home (T4), and to turn on the dryer when the washing machine completes the wash and the outside weather is bad (T5).

\textbf{T6: Modify an Automation}. Participants had to update one of the automation they created or one provided by us changing whatever they like (e.g., adding a condition, changing/adding triggers or actions). 

After completing the tasks, we asked participants a set of questions for collecting feedback regarding: i) comments about the study and the tasks, ii) aspects they considered problematic in interacting with the simulated system, iii) the things they liked most and least about the simulated system and iv) open suggestions for possible improvements.

\subsection{Participants and Collected Data}
We recruited two groups of participants, one in \Cagliari for the VR museum scenario and one in \Pisa for the AR home. The VR group consisted of 14 participants, 5 females and 9 males, aged between 19 to 33 y.o. ($\bar{x}=24.0, \sigma=4.1$). The AR group consisted instead of 15 participants, 11 females and 4 males, aged between 40 and 62 y.o. ($\bar{x}=52.0, \sigma=7.3$).
In both groups programming experience was very low. 10 VR and 7 AR participants affirmed having no prior programming experience, while the others self-rated their experience as low. 
As for users' familiarity with AR/VR, only one user in the AR group had used tools for creating an AR environment before; apart from that user, the other participants were overall unfamiliar with all types of XR applications (VR: 4 low familiarity, 10 none; AR: 9 low familiarity, 5 none). In contrast, they had more experience in using voice assistants (VR: 8 familiar, 6 unfamiliar; AR: 6 medium or good familiarity, 9 no or limited familiarity). Also, users were overall unfamiliar with systems supporting automation (AR: 11 no familiarity at all, 4 limited familiarity; VR: all participants reported no experience).

Throughout the study, we collected and analysed various types of data to gain insights into how participants structured their automation rules and interacted with the simulated system. 
In addition, we tagged the transcription of the dialogues between participants and the simulated agent to identify useful patterns in their structure (see Section~\ref{sec:formative-findings}).

\subsection{Identified Dialogue Flow}
\label{sec:formative-findings}
In both settings, users initially formulated rules that were ambiguous or incomplete, requiring refinement. The simulated conversational agent played a crucial role in assisting users by prompting clarifications or rephrasing. For example, in AR, the agent helped refine conditions such as the timing of turning lights on and off. Likewise, in VR, the chatbot asked users to specify additional details about triggers or conditions, such as identifying the artwork associated with playing an audio description. From the analysis of the dialogues between participants and the simulated agent, we identified a structured dialogue flow that aligns with the framework proposed by Mugunthan and Gibbons~\cite{mugunthan2023} in the domain of AI image generation. Thus, we adapted that framework for the domain of automations specification: the flow we propose consists of the four stages in~\cite{mugunthan2023}(Define, Explore, Refine and Export), plus an additional one we introduced (Confirm), to provide a final check on the result since the content generated is not visual (the agent asks confirmation from users before saving the automation). Below we describe the goals of each phase.

\begin{figure}[t]
    \centering
    \includegraphics[width=0.7\linewidth]{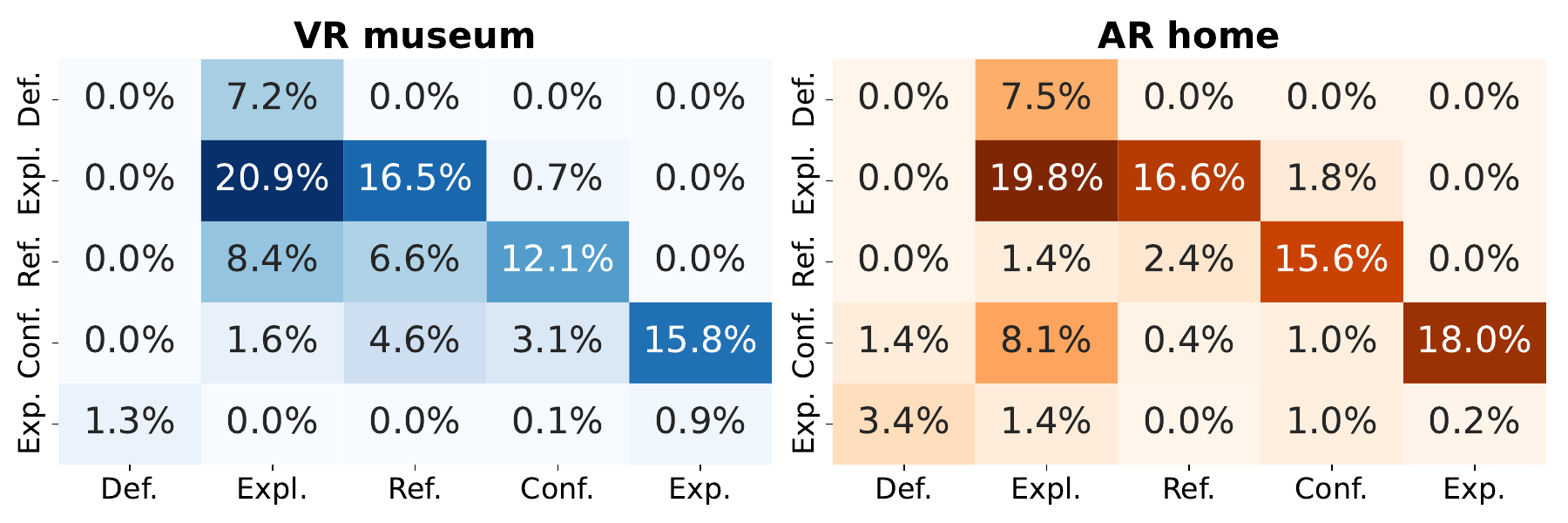}
    \caption{Transitions among the identified stages (Define, Explore, Refine, Confirm and Export) in the collected conversations during the formative study in the VR museum (left) and AR home (right). Rows are the starting stages, columns the targets. Values are normalised on the total number of conversation turns.}
    \label{fig:dialog-flow}
    \vspace{-18pt}
\end{figure}

\textbf{Define}. 
In this initial stage, users articulate high-level automation goals, in vague or abstract terms. Rather than specifying precise conditions and actions, they express general desires such as \enquote{I’d like the environment to be comfortable when I arrive}. At this point, the conversational agent plays a crucial role in helping the user refine these broad objectives into more actionable automation intentions.

\textbf{Explore}.
Once a general goal is established, users begin exploring how to implement it within the system's capabilities. This phase often involves discovering available objects, sensors or devices that can fulfil the automation goal. Users engage in a dialogue with the assistant, asking about options and available configurations. For example, in AR, if a user wants improved home security, the assistant might suggest motion sensors, smart locks, or lighting automation. In VR, users tended to interact with the objects and asked the agent which were its interaction options.

\textbf{Refine}.
After defining a preliminary rule, users refine it to ensure completeness and clarity. This includes specifying conditions (e.g., \enquote{only if the light is on}), constraints (e.g., \enquote{for 20 seconds}), or additional details (e.g., \enquote{basically here [points the floor], at a short distance from the artwork}). The assistant prompts clarifications, helping users resolve ambiguities. In AR, users iteratively refined their rules through verbal interactions, whereas in VR participants complemented refinements requests to the agent with pointing, manipulating or touching the surrounding objects.

\textbf{Confirm}.
When the dialogue converges to a point where the agent has all the required information to create an automation, it summarises the collected information and asks for a confirmation before saving the automation. After this step, users can ask to save the current definition, or iterate further based on additional considerations. 

\textbf{Export}.
This is the final stage of the process. After receiving confirmation about the understanding of the user's intent, the simulated agent stores the rule among those available in the system.

\autoref{fig:dialog-flow} summarises the analysis of the conversation flows in both settings. The heatmaps clearly show that from the \textit{Define} we usually switch to \textit{Explore} and there the conversation stays for most turns (VR: 20.9\%, AR:19.8). After that, the conversation moves to \textit{Refine} (VR:16.5\%, AR:16.6\%). In this stage, the most likely move to \textit{Confirm} (VR:12.1\%, AR:15.6\%). Another option is going back to \textit{Explore} (VR:8.4\%, AR:1.4\%), which usually happens when the participant wants to consider other options for the automation, or staying in this stage for further refinement (VR:6.6\%, AR:2.4\%). 
From the \textit{Confirm} option, the most likely move is towards \textit{Export} (VR:15.8\%, AR:18.0\%). It happens sometimes that participants were not satisfied with the summary, so they moved to \textit{Refine} (VR:4.6\%, AR:0.4\%) or \textit{Explore} (VR:1.6\%, AR:8.6\%).

\subsection{Patterns in AR and VR}

\textit{Participants split complex automations into simple ones}. In both the VR and AR experiments, users preferred to create simple automations instead of complex ones. When faced with complex scenarios requiring multiple triggers, conditions, and actions, users tended to break them down into multiple simpler rules rather than constructing a single intricate automation, a trend that is already documented in the literature~\cite{ur2014,pacpac-rules}. For instance, in the AR group, for T4 and T5 we expected, across all users, 30 complex rules in total (15 for T4 and 15 for T5) whereas we registered 47 rules (25 simple rules and 22 complex rules) . Thus, some complex rules were expressed by users splitting them in simpler rules. We registered the same pattern in the VR group, even though it was less frequent.

\textit{Diverse Expression Styles and Order of Elements}.
Participants expressed automations in varied formats, without following a dominant pattern. Some started by defining the action first, others began with the event, and a few prioritized conditions. In AR, 65 automations started with an action, 84 with an event, and 13 with a condition. A similar distribution was observed in VR (65 event, 32 action, and 7 condition), indicating that systems should be flexible enough to process multiple structuring formats without enforcing a strict syntax.



\textit{Preference for Interaction Modality}.
In AR, users clearly preferred voice interaction over simulating camera-framing selection methods, thereby finding natural language-based interaction with the agent more straightforward. Participants addressed the conversational agent in different ways: referring directly to the device, describing the result they wanted, or explicitly calling on the assistant (e.g., \enquote{Bot, turn on the light}). Instead, VR participants created automations interacting with the elements in the environment through different modalities, frequently omitting to explicitly mention the objects they referred to in speech, relying on the system to infer them from the interaction context. This indicates that we need to collect differently the context information in AR and VR.

\textit{Level of Completeness in Initial Rule Formulation}.
In VR, a slightly higher percentage of automations were immediately executable upon first formulation (78 out of 104 rules: 75\%), whereas in AR (110 out of 162: 67\%)   initial user-generated rules were more ambiguous, requiring further refinement rounds. Thus, in AR users seemed to rely more on the agent for clarification, whereas in VR, they described the automation more clearly, after they decided their intent.

\textit{Challenges in Modifying Existing Automations}.
When asked to modify existing rules, participants tended to redefine the entire automation rather than specifying the exact change. This behaviour was consistent across both environments, possibly due to the participants' unfamiliarity with automation structures. The authoring system should take into account such preference, interpreting such request as incremental modifications rather than as full rule restatements.

\textit{Modification Approaches and User Confidence}.
While both experiments show a preference for restating rules rather than specifying incremental changes, users in VR appeared to struggle more with modifications due to a \enquote{lack of imagination} on what to do differently, or concerns about the possible correctness of the interaction. In contrast, AR users seemed more confident in restructuring rules, possibly due to greater familiarity with the services offered by the home appliances, compared to possible VR interactions.

\section{Tell-XR}
\begin{figure}[t]
    \centering
    \includegraphics[width=1\linewidth]{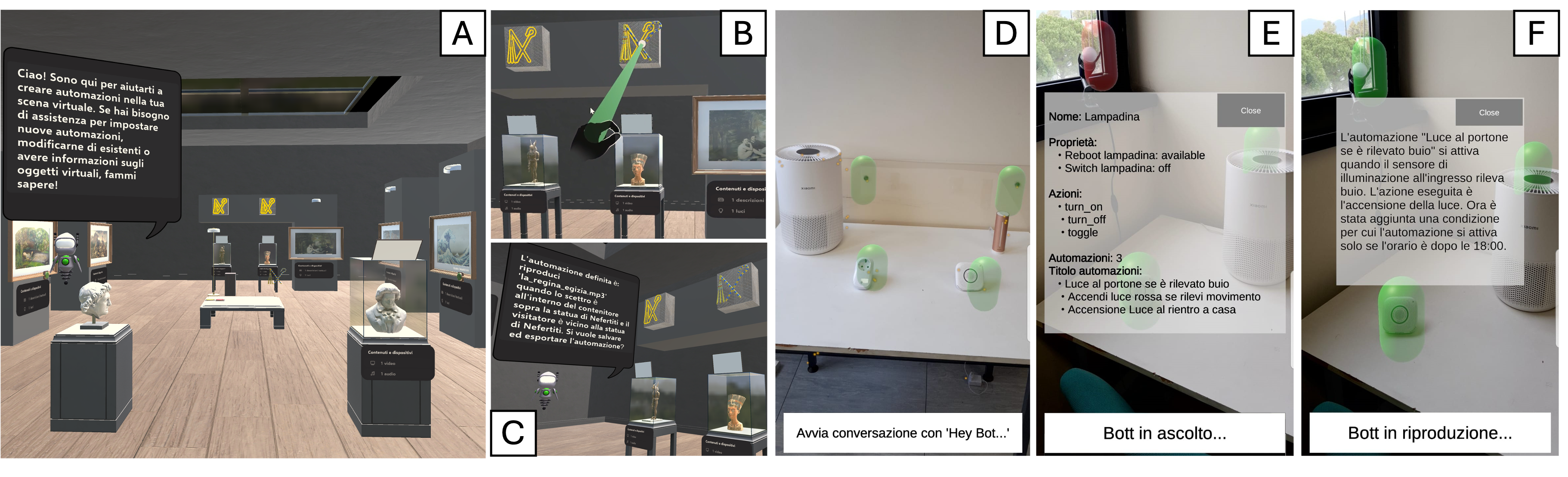} \\
    \caption{Examples of the Tell-XR interface. The VR museum environment (A), the sceptre in the socket condition (B) and confirming the example automation (C). In the AR home environment, highlighting the interactable devices (D), showing the light capabilities (E) and confirming the modified automation (F).}
    \label{fig:walkthrough}
    \vspace{-5pt}
\end{figure}

To demonstrate the Tell-XR's capabilities, we detail the creation of two examples automations in VR and AR (see \autoref{fig:walkthrough}). We report the dialogue between the bot and the user, as a shortened and adapted version of real input we collected in the evaluation (see \autoref{sec:evaluation}).

\noindent\textbf{Complex rule in VR}. Angie, a young art history teacher, wants to create an activity for her students about Egyptian art. She aims to enable them to activate an audio guide using a sceptre placed above the artworks they are interested in. She decides to create an automation with the bot. 

\vspace{0.2cm}\noindent
\angieInline{Hey Bot}
\tellxr{Hi! I'm here to help you create and manage automations in your virtual scene. If you need assistance, let me know what you want to do!}
\angie{What multimedia files are available for the statue of Nefertiti?}
\tellxr{The available multimedia files are: Audio: 
 [\dots],
Egyptian Queen.mp3, [\dots]
Video: [\dots]
If you need more information or want to create an automation, let me know!}
\angie{Yes, I want to play Egyptian Queen when the sceptre is inside the box above this statue.}
\tellxr{I can set up the automation to play the file 'Egyptian Queen.mp3' when the sceptre is inside the container above the statue of Nefertiti. Do you need to add any other conditions or events? Do you want to proceed with this automation?}
\angie{I want also the student to be near the statue of Nefertiti.}
\tellxr{So the automation will be: play 'Egyptian Queen.mp3' when the sceptre is inside the container above the statue of Nefertiti and the visitor is near the statue of Nefertiti. Do you want to add any other conditions or events, or can we proceed with this automation?}
\angieInline{That's fine.}
\tellxr{The automation you defined is: [\textit{repeats automation}].
Do you want to save and export it?}
\angieInline{Yes.}
\tellxrInline{The automation has been saved.}

%

\textbf{Modifying a rule in AR}. 
Bob is a businessman who regularly travels between his home and place of work located in another town. He wants his entrance light to turn on when he nears home for security. He sets up an automation with an intelligent assistant but notices the light sometimes turns on unnecessarily. To save energy, he modifies the rule to activate only in darkness.


\noindent
\bobInline{Hey Bot.}
\tellxr{Hi! How can I help you with your smart home today? Do you want to create an automation?}
\bob{No, I want to modify an automation.}
\tellxr{Which automation would you like to modify? Can you provide some details?}
\bob{The lighting management for the front door.}
\tellxr{I found several automations for the entrance lighting. Do you want to modify the activation upon return or when presence is detected?}
\bob{The activation upon return.}
\tellxr{Would you like to add conditions, such as a time range for activation upon return?}
\bob{No, only when it's dark.}
\tellxr{The automation will turn on the entrance light when the GPS sensor detects that the user is near home, but only if it is dark. Do you want to save the automation?}
\bobInline{Yes, I do.}
\tellxrInline{The automation has been updated to turn on the entrance light upon return only if it is dark.}

\subsection{Technical implementation}
\begin{figure}[t]
    \centering
    \includegraphics[width=1\linewidth]{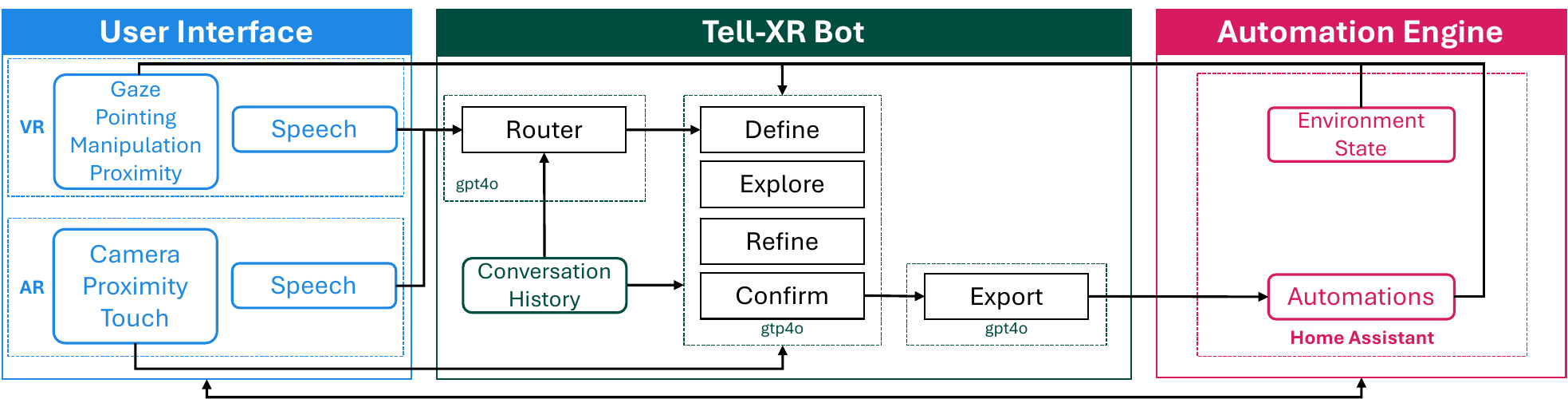}
    \caption{The Tell-XR architecture consists of three components, 1) the User Interface (light blue), which provides the multimodal input, 2) the Automation Engine, currently implemented on top of Home Assistant~\cite{home-assistant} and 3) the Tell-XR bot, consisting of several sub-modules to handle the phases in the conversation identified in Section~\ref{sec:formative-findings}}
    \label{fig:bot-architecture}
    \vspace{-16pt}
\end{figure}

Figure~\ref{fig:bot-architecture} shows the overall architecture of the Tell-XR system, which consists of three components, separating the different concerns of the solution. 

\textbf{User Interface (UI)}. This is the layer supporting the interaction with the end-user. We have two different implementations, one for VR and one for AR. They support the environment presentation, synchronise the view with the Automation Engine, and contain modules for collecting user input. The speech is the main channel, as the end-user verbalises their intent, asking the bot to create automations for them. This module is also responsible for converting user voice commands into text and transforming the textual responses generated by the bot into audio. Specifically, the VR scenario uses OpenAI Whisper \cite{openai2022whisper} for speech-to-text conversion and OpenAI tts-1 \cite{openai2023tts} for generating audio responses, while AR exploits Azure AI Speech  \cite{azure2025}  for both TTS and STT. 
As Section~\ref{sec:formative-findings} explains, users complement speech with information expressed through different modalities. The UI component contains a set of submodules that monitor the other available sources, filter the information, and serialise the relevant data in a textual format. The Tell-XR bot can use this context to interpret the user's intention.

Both implementations provide: i) the list of all virtual/physical objects currently present in the environment and the properties and services that characterize them, ii) the list of currently framed objects (smartphone camera in AR or view frustum in VR), and iii) the list of configured automations. 
In VR, the input streams include head orientation and visible objects, user position, proximity to objects, information about pointed and grabbed items, available media files, distance and direction of a given object relative to others.
In AR, we use a position-based solution to anchor digital elements in the physical space. Users place icons by tapping the screen, selecting devices from a menu, and saving their positions for persistent display upon future scans. The VR version exploits Unity and MRTK 3.0~\cite{mrtk}, while the AR exploits Unity, AR Foundation and AR Core.


\textbf{Automation Engine}. 
This component manages the relevant state of the XR environment, encompassing virtual objects and physical devices. We leverage Home Assistant~\cite{home-assistant}, an open-source home automation platform, to store and execute various automations. Home Assistant is an open-source home automation platform that allows users to control and automate smart home devices from diverse manufacturers through a wide range of integrations.  To include the virtual objects and content we developed an extensible taxonomy of VR objects that outlines their state variables and operations for automation as in~\cite{ecarules4all}. In VR, by navigating the Unity scene graph, we generate device instances that reflect the state of virtual objects within the platform. In AR, physical objects are associated to their corresponding representations during an initial configuration phase. The automation services incorporate the necessary code for Home Assistant to manage automations and synchronize virtual object visualizations in Unity. Automations are stored in a JSON file, containing triggers, conditions, and actions, with Tell-XR programmatically adding user-defined automations and retrieving the current configurations.

\textbf{Tell-XR Bot}. This component is the core of the authoring system. It is responsible for assessing the users' utterances, interpreting their intentions, and guiding them until the desired automation is clear enough to be stored in the system. 
We built the bot in TypeScript, using LangGraph~\cite{langchain} to prompt the LLM GPT-4o~\cite{openai2024gpt4o}. 
The component is an LLM-based chatbot exploiting Retrieval-Augmented Generation~\cite{lewis2020} containing different nodes.
It is an AI multi-agent structured as a graph with cycles, where each node represents a phase in the dialogue flow. It uses a \enquote{Human in the Loop} approach, waiting for user input at breakpoints to clarify previous dialogues and decide the next step. Special nodes represent specific operations, and when a request is received, it follows a path in the graph, using external tools via ToolNode to gather information or perform actions, thereby enhancing its response. Communication among modules is achieved using REST APIs over HTTP.
The entry point node is the \textit{Router}, which receives the Speech-to-Text from the UI component. By considering the current request and conversation history, it forwards the request to one of the four nodes dedicated to each of the automation definition phases discussed in Section~\ref{sec:formative-findings}. Such a structure guarantees that users can jump from one phase to the other following their thoughts, without predefined ordering.  

The \textit{Define, Explore, Refine}, and \textit{Confirm} nodes receive as context the full spectrum of information available in the system (see Figure \ref{fig:bot-architecture})  including 1) the conversation history, 2) the speech-to-text input, 3) the filtered textual representation of all the other input modalities, 4) a textual representation of all the objects in the environment and 5) the list of currently defined automations. We clear the conversation history each time the automation is exported to the automation engine. Each node has also a different prompt that guides the bot response generation adapted for the AR and VR scenarios (available in the additional material), which is returned to the UI component for the presentation. 
The bot uses Role-based, Instruction-based, and Few-Shot Prompting to enhance its responses. Role-based Prompting assigns a specific function to the model for better context, while Instruction-based Prompting provides clear guidelines on how to respond. Few-Shot Prompting includes examples to improve generalization and accuracy.
When the end-user agrees with the automation proposed by the bot after a certain number of conversation turns, the confirmation module passes the collected information to the \textit{Export} submodule, which generates a JSON description of the identified automation and stores it among those in the environment. After this step, the environment is ready to trigger the automation.
Also, from the history of the conversation with the user, the system can understand whether the user wants to create a new automation or to modify an existing one: in the latter case, the list of existing automations needs first to be retrieved to identify the automation to change.
We used three different instances of the same LLM for implementing this component: one dedicated to the \textit{Router} module, one for the \textit{Export} module, and the last one for the modules dedicated to user's dialogue. We did this because we have empirically found that for the transformation between the intent mediated by the chatbot and the JSON version of the rule, it is better to have a dedicated LLM with a prompt specifically designed for the task. This helps to avoid hallucinations and formatting errors in the rule.

\section{Evaluation}
\label{sec:evaluation}

We conducted a usability test similar to the WoZ study, involving two user groups who interacted with a prototype in two scenarios: a museum (VR) and a smart home (AR). Both groups performed the same tasks and filled out questionnaires to assess workload~\cite{hart1988}, user experience~\cite{ueq-s}, the conversational agent~\cite{bus11}, and to provide more general suggestions. Also, we aimed to evaluate how well the users' rules fulfilled the tasks,  and gather data on task time.

\subsection{Participants}
\hspace{16pt}\textbf{VR Group}. 12 participants (6 women, 6 men), with age ranging between 16 and 43 ($\bar{x}=26.2, \sigma=10.0$) participated in the study. We recruited them by distributing invitations via mailing lists of individuals and groups we collaborate with (not including people involved in this work). In particular, we reached out to students from a high school with which \Cagliari has an ongoing collaboration. All participants are students or workers in humanities or students enrolled in a course involving activities for promoting cultural heritage. 

\textbf{AR Group}. 13 participants (10 women, 3 men) with age ranging between 30 and 64 ($\bar{x}=48.2, \sigma =11.1$) were involved in the study. They were recruited through messages circulated on mailing lists of Institute-2 and of the larger research area which it belongs to. Participants were either workers in administrative roles, or researchers working in fields not related to computer science. In any case, no participant was involved in our research and development work.

\textbf{General Information}. Before the test, we collected information about participants’ prior experience. Using a 1-5 scale (1= no familiarity; 5=high familiarity), we assessed their familiarity with voice assistants (e.g., Siri, Alexa) and conversational AI (e.g., ChatGPT). Among AR participants, 5 rated their experience as ‘1’, 3 as ‘2’, and 5 as ‘3’. In the VR group, 7 rated it as ‘3’ and 2 as ‘4’. Regarding specific assistants, 4 VR participants knew Siri, 5 Alexa, 1 Cortana, 5 ChatGPT, and 1 Google Assistant, while in AR, 7 knew Alexa, 2 ChatGPT, 2 Siri, and 1 user Google Assistant.
Familiarity with XR applications was low. In the VR group, 8 participants rated their knowledge as ‘1’, 3 as ‘2’, and 1 as ‘3’, with none able to name a VR application, only headsets. In the AR group, 12 reported no familiarity (‘1’), while 1 rated it as ‘2’, mentioning the IKEA app and AR experiences in museums.
Regarding automation systems, only one VR participant rated their familiarity as ‘4’, mentioning Google Home and Samsung SmartThings, while all others rated it as ‘1’. In the AR group, 12 rated it as ‘1’, and 1 as ‘2’, mentioning the TaDo system for smart heating control.
Participants had no programming experience except for one in each group, both with limited HTML knowledge. No compensation was provided for participation.

\subsection{Procedure}
Users that accepted to participate to the test received by email (AR) or on paper (VR) an introduction to the study, describing its motivations and goals, as well as some high-level description of the application and its main functionalities. Also, they received a consent form in which the main goal of the study was described, the advantages and disadvantages in participating in it, and how the personal data were dealt with. To be involved in the study, users had to fill in the consent form and sign it. For underage participants, consent was obtained from both the participants themselves and their parents or legal guardians.

Participants in the VR group completed the test either in our laboratory or in the facilities of the high school. They used an Oculus Quest 3 or 3S headset to interact with the virtual environment, including a fictional museum room containing paintings, sculptures and other material for enhancing the visit, such as unexhibited work, textual descriptions, and multimedia material. 
For the AR setting, participants come to the lab for the test. The office in which the test had to be carried out was equipped with devices that can be found in smart home scenarios (e.g. lamp, air purifier, motion sensor, presence sensor). For the test, they used a Samsung Galaxy S10 smartphone.

The test started with a familiarisation phase, in which participants were requested to: i) Ask the assistant to introduce itself to the user and tell the user how it handles privacy, and what it can do; ii) Familiarize with the objects in the room, to understand their status and the automations associated with them; iii) Create an automation of their choice, by using the application. In the VR setting we also included a small video showing how to interact with the environment in VR, since we expected a low familiarity with VR applications. 

After that, users were provided with the description of the scenarios/tasks to solve, using a paper form (VR), or a web page in which both the tasks to solve and the questions to answer were available (AR). We submitted to users six different tasks, covering not only the creation of automations (5 tasks) but also change of automations (1 task). In addition, the tasks were submitted with an increased difficulty level:

\textbf{T1} and \textbf{T2} asked to create a \textbf{simple automation} (1 event + 1 action). In VR, T1 asked to automate spotlights to enhance the visibility of a painting when visitors approach. T2 requested to display an information panel about an artwork when the visitor presses a button.
In AR, T1 asked to create an automation to prevent kitchen hazards, such as those connected with e.g. forgetting a pot on the stove, while T2 asked to automate the front door lighting to ensure that it is on when the user arrives home in the dark.
\textbf{T3} asked the user to \textbf{invent an automation} of a user's choice. 
\textbf{T4} and \textbf{T5} requested to create a \textbf{complex automation}, including conditions and one or multiple actions. In VR, T4 asked to unlock hidden information about an artwork when a visitor places a related historical object in a special container, while T5 to play multimedia content when the visitor approaches Beethoven’s statue, but only if multimedia playback is enabled in the room. In AR, T4 asked to automate the air purifier to stay on while the user is in the living room, while T5 required to create an automation to regulate temperature for houseplants by operating a window when it is too hot on weekends.
\textbf{T6} asked the user to \textbf{modify} the (simple) rule associated with T2, to obtain a complex rule (by adding a condition).

After each task, users filled in a NASA-TLX~\cite{hart1988} questionnaire, to evaluate the effort requested to carry out that task. After evaluating the last task, users rated their overall user experience with the system by filling in the UEQ-short~\cite{ueq-s}, then filled in the BUS-11~\cite{bus11} questionnaire to assess the conversational agent. Then, they further provide observations on positive/negative aspects of the system, as well as recommendations for possible improvements.

\subsection{Results}
\begin{figure}[t!]
    \centering
    \includegraphics[width=\textwidth]{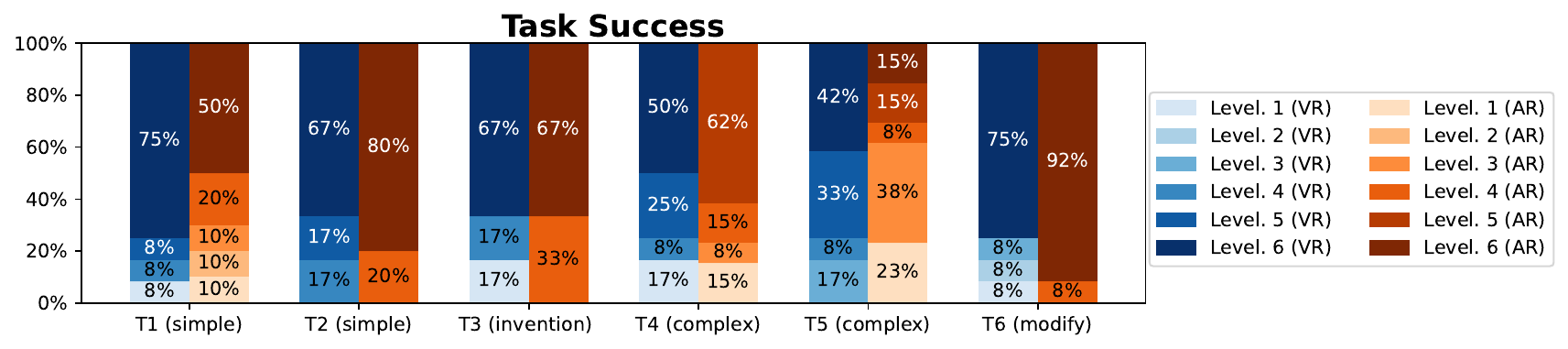} \\
    \includegraphics[width=\textwidth]{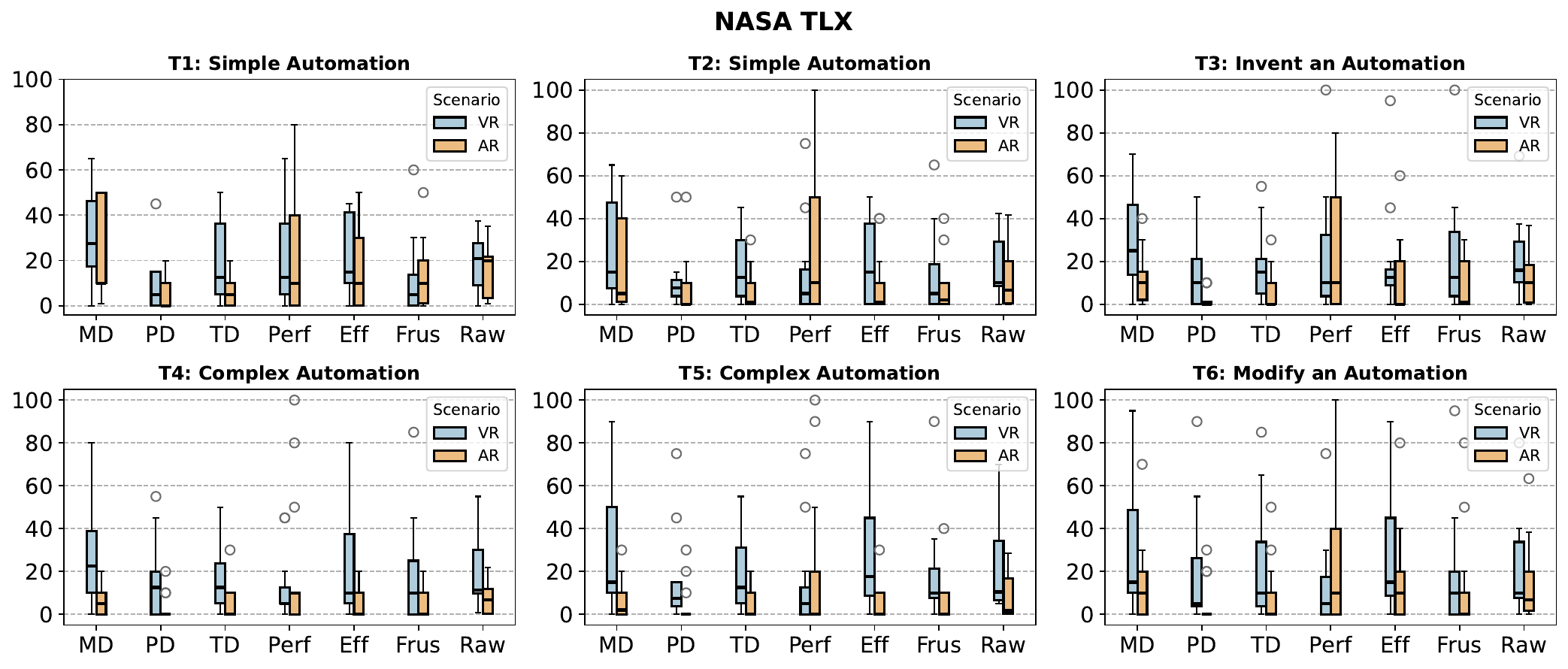} \\
    \includegraphics[width=\textwidth]{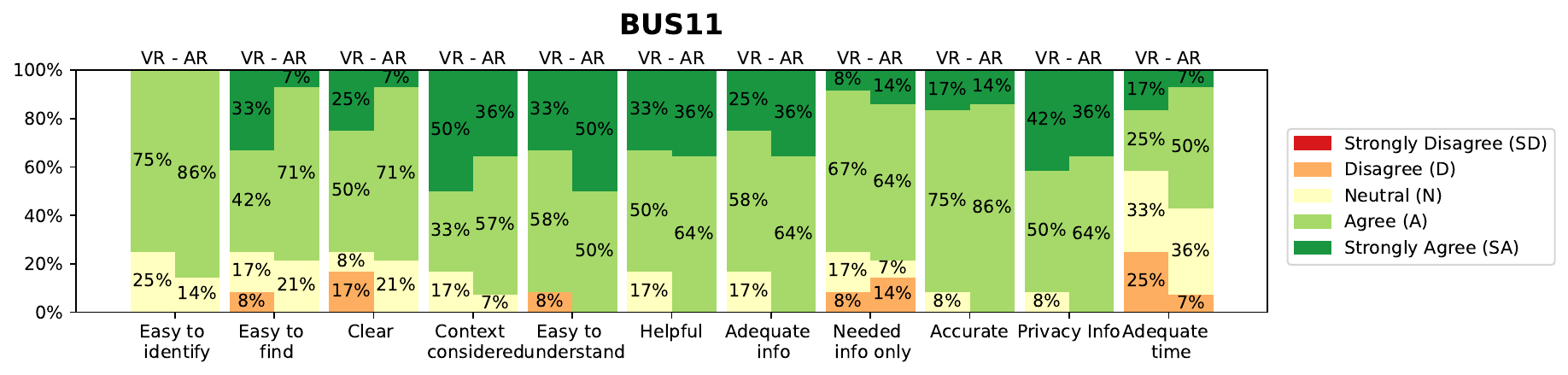} \\
    \includegraphics[width=0.63\textwidth]{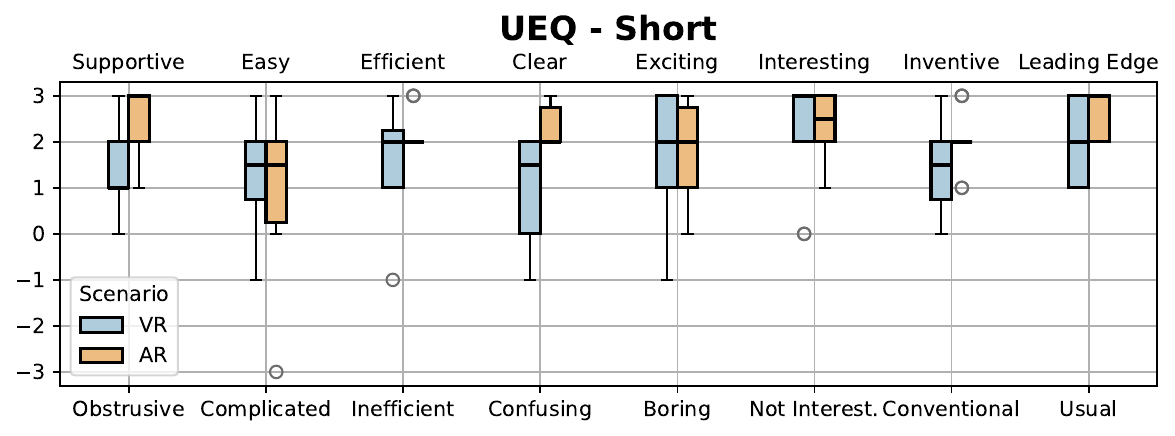}
    \includegraphics[width=0.34\textwidth]{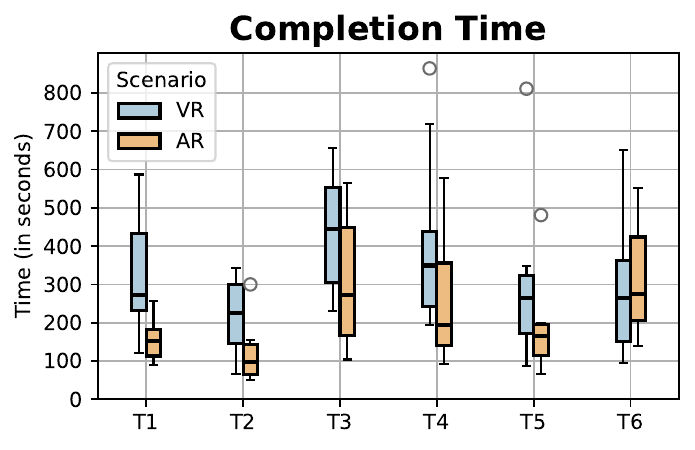}
    \caption{Summary of the quantitative data collected during the user test through  (from top to bottom) NASA-TLX~\cite{hart1988}, BUS-11~\cite{bus11}, UEQ-short~\cite{ueq-s} questionnaires and task completion time. For each dimension we report the VR values on the left (light-blue in box-plots) and the AR values on the right (orange).}
    \label{fig:evaluation}
    \vspace{-20pt}
\end{figure}

Figure~\ref{fig:evaluation} summarises the quantitative data for the AR and VR groups. 

\textbf{Task Success}. First of all, we analysed the success level in defining the automation for the considered task.  
We consider three levels of success for
each side, the user ($U$) and the bot ($B$): correct ($c$), partially correct ($p$), and wrong ($w$). The agent correctness was assessed according to what extent it fulfilled the request made by the user. Instead, the users' correctness depended on the extent their intent was in line with what the task requested. Thus, a partially correct agent occurred when it failed at setting one element of an automation (either the trigger, the action or the condition, if present). If more than one rule element was wrongly set by the agent, the agent was classified as wrong. We give more relevance to the agent error (compared to user error), finally grouping the various cases into the following 6 levels (higher the score, better the evaluation): 1) $U_wB_w$ | $U_pB_w$ | $U_cB_w$, 2) $U_wB_p$ | $U_pB_p$ 3) $U_wB_c$, 4) $U_cB_p$ 5) $U_pB_c$ 6) $U_cB_c$.

Overall, we had a good success rate across all tasks. In VR tasks T1, T2, and T5 performed well. One main issue with the bot was that it sometimes hallucinated triggers or misunderstood actions as triggers. After a few interactions, the bot would often guess the virtual objects involved in automation without asking the user for confirmation. In T3, participants struggled to define their goals. Two had partially correct definitions, leading to a wrong automation (level 1), while the other two were guided correctly to the right automation (level 4). 
In T6, we again faced challenges with modifying automation. Two participants started with incorrect requests; the bot understood one correctly (level 3) but failed on the other (level 1). Another participant asked for an incomplete modification, and while the bot suggested a correct change, it eventually hallucinated.

In the AR setting most user errors raised from participants incorrectly responding to the agent, leading to agent errors. Notably, there were instances of hallucination, such as the agent suggesting a non-existent heat sensor in the smart home context considered. In T2, similar errors occurred, though less frequently, primarily involving users answering ambiguously. Also, a user failed to confirm a rule, resulting in a task failure. In T3, users sometimes provided unhelpful responses to the chatbot’s requests. T4 saw an increase in user errors as many confirmed simple rules whereas a complex automation was necessary. In T5, users frequently confirmed incomplete rules, and inconsistencies led to further chatbot errors. In T6, ambiguous responses caused chatbot errors and one hallucination. Two users chose not to modify rules they identified, leading to task failures: in one case the user realised that the identified rule did not correspond to the user's intended one, in the other case the user mistakenly viewed a modification as a deletion plus the addition of a new rule and he deleted a rule that was not the one required by the task.

\textbf{Task time}. Task completion time is higher in VR for all tasks but T6. This was expected, considering the effort required to navigate and explore the virtual environment.

\textbf{Nasa-TLX}. In the AR scenario, all the values reported by users were low. The only (slight) peak we observed was in T1, which got the highest overall raw value ($Q_2= 20, \Delta Q= 18.33$). However, that value is still low, showing that the system did not put a significant workload on users while interacting with it. We expected a slightly higher workload on T1 since it is the user's first exposure to a novel system. In particular, in the Mental Demand dimension, we registered decrements in case of a second exposure to the same type of task (a decrement occurred in T2, and another one occurred in T5), which suggests that users perceived a less cognitive demand when they had to carry out a type of task they were already confronted with before.  
In VR, we expected an increase in workload due to immersion and discomfort from the headset. However, average scores across all dimensions were below the reference values reported in previous research~\cite{morten2021}, indicating that the system did not impose a significant workload. The maximum value we registered is on the Mental Demand dimensions for T1 ($Q_2=27.5, \Delta Q=28.8$), which could be explained considering that participants required to get familiar with the environment in the first task. The ‘Physical Demand’  increased steadily until T4 and then decreased, suggesting initial fatigue but eventual adaptation. We also observed a peak in all dimensions in T3, which indicated increased effort when users were not given an overall automation goal.

\textbf{BUS-11}. The stacked bar chart in  \autoref{fig:evaluation}  shows, for each aspect, the percentage of users that provided each agreement level to the various statements of BUS-11. In the AR setting, 2 users expressed disagreement and 1 was neutral about the fact that the chatbot provided only the needed information; also, 1 user disagreed and 4 users were neutral about the chatbot's responsiveness. 
In VR, we registered a similar positive trend. Points for improvement are time response (3 disagreed and 4 neutral), which mostly depended on network latency, and clearness of communication with the bot (2 disagreements). 

\textbf{UEQ-S}. This questionnarie~\cite{ueq-s} has 8 items, each consisting of a pair of negative and positive terms. Participants rated each item on a 7-point scale, from -3 (fully agree with the negative term) to +3 (fully agree with the positive term). On average, the system received positive scores for the pragmatic (AR: 2.06, VR: 1.33) and for the hedonic (AR: 2.21, VR: 1.85) qualities. This suggests that the system performs well in terms of both pragmatic and hedonic qualities, resulting in an overall positive user experience. 
In particular, AR users on average perceived the system to be leading edge ($Q_2 = 3, \Delta Q = 1$), supportive ($Q_2 = 3, \Delta Q = 1$). Thus, based on the data collected, the user experience of the system was overall very good. In VR, users perceived the system as leading edge ($Q_2=2.0, \Delta Q=2$) and interesting ($Q_2 = 3, \Delta Q = 1$).

\textbf{Open Questions}.
\textit{State the three most positive aspects of the system}.
Users highlighted several positive aspects of the app across both AR and VR settings. In AR, 7 out of 13 found it easy to navigate and intuitive, requiring no special skills. Participants appreciated the clear responses (6 out of 13) and the accuracy of the information provided. The app's flexibility and automation capabilities, including voice commands and device control, were well-received. Notably, two users found it valuable to get information about devices by framing them with the camera.
In the VR setting, 4 out of 12 praised the ease of creating automations via voice commands, while 5 appreciated the chatbot's clarity. Users enjoyed the \enquote{\textit{modern}}, interactive VR environment and the ability to explore, describing the application as \enquote{\textit{fun, innovative, and engaging}}.

\textit{State the three most negative aspects of the system.}
%
In AR, two users had nothing to report, while 11 mentioned app delays and 7 difficulties with voice recognition, including the need for clear articulation and problems understanding short responses. Also, some users experienced repeated prompts and felt that the choice of options was limited. One user noted that \enquote{\textit{If a desired option was not among those proposed, it takes some time to reset the dialogue to get what I want}}, and another was unsure on whether to act or wait for feedback. 
In VR, 5 out of 12 participants reported no significant issues, 4  experienced delays in chatbot responses, with one stating, \enquote{\textit{the bot sometimes takes a while to respond}}. Three cited usability concerns, like the bot interface being occluded and issues with image focus in the headset. Suggestions included enhancing interaction variety and addressing minor graphical and performance issues.

\subsection{Discussion}
\label{sec:Discussion}

\textbf{Generalization Potential}. 
The findings from the two user studies indicate that the Tell-XR approach has significant potential for generalization across various XR interaction settings. Although conducted in controlled environments, the studies show that end-users without programming skills can successfully automate different XR contexts. A key innovation of the system is its use of an LLM-based intelligent agent to assist users in creating and modifying event-condition-action rules for automations in XR. 
The evaluation demonstrate that
this approach successfully worked across various domains, user goals and interaction paradigms, whether using head-mounted displays or smartphones.
%
%
Moreover, the approach was equally effective for automating both digital objects in VR, whose behaviour could differ from the real world (e.g., objects can disappear, fly, etc.) and in AR, which considered more constrained physical devices.  
%
The Tell-XR multimodal architecture integrates visual and auditory inputs, making it flexible for both entertainment-focused (virtual museum) and utilitarian applications (smart home). This suggests that the approach could be easily extended to other XR environments (e.g., Mixed Reality) and domains (e.g., education). 


\textbf{AR and VR: Similarities and Differences}. 
The user study highlighted  some similarities in experiencing Tell-XR in AR and VR. We registered a good task success rate in both settings, even though some difficulties persists in the creation of complex rules. The most frequent causes are incorrect requests by users and hallucinations from the bot. Both environments had relatively low reported workload scores~\cite{hart1988} and similar patterns through the test, with the first task (T1) showing a slightly higher mental demand, likely for user adaptation. Subsequent tasks, especially when users encountered similar tasks again (e.g., T2 and T5), resulted in a decrease in perceived cognitive demand. This suggests that users were able to adapt to the system quickly, reducing cognitive load as they became more familiar with the task types. Participants in both AR and VR rated the system positively for pragmatic and hedonic qualities, indicating a generally good user experience. Users in both environments appreciated the system as being innovative and engaging, with room for improvement in response time and clarity of communication with the bot. 

We registered also interesting differences between VR and AR. In VR, user had more difficulties in defining their goal when it was open-ended (T3). This is related with a lack of familiarity with VR interactions, which we must take into account for improving the bot. In addition, wearing the headset caused a particular pattern in the perceived physical demand that increased gradually until T4 and then decreased, suggesting that initial physical discomfort was mitigated as users adapted to the VR setup. In AR, the smartphone-based interface and the more familiar physical context led to a more straightforward interaction. Here, the main source of problems in task completion are hallucinations, such as suggesting non-existing sensors or failing in confirming the rules.

\textbf{Guiding Users in Modifying Rules.} 
In the formative study, participants tended to redefine entire automation rules instead of specifying desired changes, indicating an 'abstraction' gap in understanding rule structure. This highlighted the need for an authoring system that can translate whole rule restatements into incremental modifications.
In the usability study, particularly in the AR setting, users demonstrated improved understanding of rule components and successfully modified rules by referencing specific devices or effects, aided by agent guidance. 
However, the VR setting was more challenging. While most users modified automations successfully, some reverted to restating entire rules, and there were instances of miscommunication with the bot. 
These findings emphasize the need for the bot to enhance its understanding and guidance capabilities for users. Future system iterations should address communication breakdowns and provide clearer feedback, enabling users to articulate incremental changes more effectively across various XR environments.


\textbf{Challenges and Limitations}. While the system satisfactorily guided users to create and modify automations, in some instances, the agent’s responses deviated from user requests because of the hallucination phenomenon, providing unexpected results to the users. The cause of such responses are various, including ambiguities in user input or limitations in the system’s ability to fully understand nuanced requests from users. This suggests that, while the system has been shown promising, its capability to effectively manage the wide range of inputs it may encounter in dynamic XR environments represents an area for further improvement. One limitation of the studies is the limited sample size. Future work could address this by expanding the number and variety of participants  to enhance the generalisability of the results.

\section{Conclusion and Future Work}
In this paper, we presented Tell-XR, an LLM-based intelligent agent able to guide end-users without programming skills in defining rule-based automations in XR environments. Through a formative study, we identified the dialogue stages through a formative study, guiding the bot's technical implementation. The usability test results show the effectiveness of the approach across different XR interaction settings and application domains: a VR museum and an AR smart home. Future work will focus on extending the approach towards Mixed Reality (headset-based AR) and improving the quality of the bot guidance, especially for modifying existing rules and for robustness to hallucinations. 

\section{Acknowledgements}
This work has been supported by the Italian PRIN 2022 \enquote{EUD4XR: End-User Development for eXtended Reality} funded by the Italian MUR and European Union - NextGenerationEU under grant PRIN 2022 EUD4XR (Grant F53D23004380006) \url{https://prin.unica.it/eud4xr/}. 

\begingroup
\renewcommand{\baselinestretch}{0.9}
\renewcommand{\normalsize}{\fontsize{8}{10}\selectfont}
\bibliographystyle{splncs04}
\bibliography{bib}
\endgroup

\end{document}